\documentclass[
superscriptaddress,
preprint,
amsmath,amssymb,
aps,
prx,
]{revtex4-2}

\usepackage{graphicx}\usepackage{dcolumn}\usepackage{bm}\usepackage{hyperref}\usepackage[normalem]{ulem}
\usepackage{braket}
\usepackage{placeins}

\usepackage{xcolor}

\definecolor{green2}{RGB}{80,205,80}

\begin{document}

\title{Fractional focusing peaks and collective dynamics in two-dimensional Fermi liquids}

\author{Adbhut Gupta}
\thanks{These two authors contributed equally.}
\affiliation{Department of Electrical Engineering, Princeton University, Princeton, New Jersey 08544, USA}

\author{Gitansh Kataria}
\thanks{These two authors contributed equally.}
\affiliation{Bradley Department of Electrical and Computer Engineering, Virginia Tech, Blacksburg, Virginia 24061, USA}

\author{Mani Chandra}
\affiliation{Department of Materials Science and Engineering, Rensselaer Polytechnic Institute, Troy, New York 12180, USA}

\author{Siddhardh C. Morampudi}
\affiliation{Center for Theoretical Physics, Massachusetts Institute of Technology, Cambridge, MA 02139, USA}

\author{Saeed Fallahi}
\affiliation{Department of Physics and Astronomy, Purdue University, West Lafayette, Indiana 47907, USA}
\affiliation{Birck Nanotechnology Center, Purdue University, West Lafayette, Indiana 47907, USA} 

\author{Geoff C. Gardner}
\affiliation{Birck Nanotechnology Center, Purdue University, West Lafayette, Indiana 47907, USA} 

\author{Michael J. Manfra}
\affiliation{Department of Physics and Astronomy, Purdue University, West Lafayette, Indiana 47907, USA}
\affiliation{Birck Nanotechnology Center, Purdue University, West Lafayette, Indiana 47907, USA} 
\affiliation{School of Electrical and Computer Engineering, Purdue University, West Lafayette, Indiana 47907, USA} 
\affiliation{School of Materials Engineering, Purdue University, West Lafayette, Indiana 47907, USA} 

\author{Ravishankar Sundararaman}
\email{sundar@rpi.edu}
\affiliation{Department of Materials Science and Engineering, Rensselaer Polytechnic Institute, Troy, New York 12180, USA}

\author{Jean J. Heremans}
\email{heremans@vt.edu}
\affiliation{Department of Physics, Virginia Tech, Blacksburg, Virginia 24061, USA}

\begin{abstract}

Carrier transport in materials is often diffusive due to momentum-relaxing scattering with phonons and defects. Suppression of momentum-relaxing scattering can lead to the ballistic and hydrodynamic transport regimes, wherein complex non-Ohmic current flow patterns, including current vortices, can emerge. In the ballistic regime addressed here, transverse magnetic focusing is habitually understood in a familiar single-particle picture of carriers injected from a source, following ballistic cyclotron orbits and reaching a detector. We report on a distinctive nonlocal magnetoresistance phenomenon exclusive to fermions, in an enclosed mesoscopic geometry wherein transverse focusing magnetoresistance peaks also occur at values of the cyclotron diameter that are incommensurate with the distance between the source and detector. In low-temperature experiments and simulations using GaAs/AlGaAs heterostructures with high electron mobility, we show that the peaks occur independently of the location of the detector, and only depend on the source-drain separation. We reproduce the experimental findings using simulations of ballistic transport in both semiclassical and quantum-coherent transport models. The periodicity of magnetic field at which the peaks occur is matched to the lithographically defined device scale. It is found that, unlike in transverse magnetic focusing, the magnetoresistance structure cannot be attributed to any set of ordered single-particle trajectories but instead requires accounting for the collective dynamics of the fermion distribution and of all particle trajectories. The magnetoresistance is further associated with current flow vorticity, a collective phenomenon. 

\end{abstract}

\maketitle

\section{Introduction}

Ballistic carrier transport in solid-state systems occurs when charge carriers scatter predominantly against the device boundaries, rather than dissipating system momentum to the lattice or undergoing mainly electron-electron scattering \cite{gupta2021PRL, gupta2021NatComm, SpectorSS1990, Heremans2004, Gilbertson2011, Lee2016, Banszerus2016}. Ballistic transport phenomena are often presented in terms of trajectories of single non-interacting particles, such as cyclotron orbits under an applied magnetic field $B$.  Here, we instead advance the conceptual picture of collective dynamics in ballistic transport \cite{gupta2021PRL, gupta2021NatComm, chandra2019, chandra-arXiv2019, shytov, bandurin2018, Khachatur2021}. We present a previously unreported periodic magnetoresistance structure occurring at low $B$, in a confined transverse magnetic focusing (TMF) mesoscopic geometry in a high-mobility two-dimensional electron system (2DES) in a GaAs/AlGaAs heterostructure. We evince that the origin of the magnetoresistance structure cannot be traced to an ordered set of single-particle cyclotron orbits, and requires consideration of \emph{all} trajectories in the device and of the collective dynamics of a fermion distribution including electron and hole excitations. Using experimental data and high-resolution kinetic simulations based on two complementary numerical schemes, we examine the factors that affect the magnetoresistance structure and lay out the essential conditions required for its manifestation. We show how the magnetoresistance correlates with current vorticity and backflow, collective properties significantly more complex than single-particle trajectories as well as harder to access experimentally. 

In contrast to ballistic carrier transport, the more common diffusive transport in the solid state, governed by Ohm's law, occurs when carrier mobility mean-free paths $\ell$ are limited by momentum-relaxing processes such as electron-phonon and electron-defect scattering which act to transfer momentum from the charge carriers to the lattice. Diffusive transport breaks down when this scattering becomes sufficiently weak and hence $\ell$ sufficiently long, as in two-dimensional particle systems in III-V heterostructures \cite{gupta2021PRL, gupta2021NatComm, Houten1989, SpectorSS1990, Heremans2004, Gilbertson2011, Molenkamp1994, deJong1995, Heremans1992, Heremans1999, Chen2005, govorov, Keser2021, Levin2018, Gusev2018, Braem2018}, in graphene \cite{SongNatComm2016, Taychatanapat, Mayorov2011, Lucas2018, Ilani2019, Lee2016, Banszerus2016, Pezzini2DMat, kumarsuperballistic, bandurin2016, bandurin2018, Berdyugin2020, Jayich2022}, in WTe$_2$ \cite{Zfei2017, Vool2021, ZZhu2015, Aharon-Steinberg2022}, or in delafossites such as (Pd/Pt)CoO$_2$ \cite{Hicks2012, Bachmann2019, Putzke2020, Bachmann2022, Moll2016, Nandi2018, Harada2021, Mackenzie2017}. Not affecting total system momentum or carrier mobility $\mu$ or $\ell$, momentum-conserving electron-electron scattering also exists, and at long $\ell$ it determines whether transport is ballistic or hydrodynamic. The ballistic regime occurs when electron-electron scattering is weak, the hydrodynamic regime occurs when it is strong, resulting in charge flow akin to a fluid. The hydrodynamic regime exemplifies collective transport and can lead to current vortex formation due to viscous drag that arises due to interactions \cite{Aharon-Steinberg2022, gupta2021PRL, Danz2D-2020, bandurin2018, Levin2018, bandurin2016}. Vorticity in the ballistic regime is on the other hand surprising because there it occurs in the absence of interactions. Recent results in ballistic transport have challenged the single-particle trajectory notion which e.g. has been used to design ballistic devices \cite{SongNatComm2016, MargalaSSE2011}. It has recently been shown that ballistic transport can produce complex current flow patterns that defy the single-particle trajectory notion and can lead to collective phenomena such as current vortices \cite{gupta2021PRL, gupta2021NatComm, chandra-arXiv2019}. In this work, the experimental magnetoresistive phenomena find an origin in such collective ballistic dynamics. These are depicted in terms of current streamlines and voltage contour plots obtained from the kinetic simulations, further compared to a quantum-coherent transport model. In related context, other work has studied voltage and charge contour plots in mesoscopic geometries \cite{EChatzikyriakou-PRR2022, Gpercebois-PRB2021}, or imaged the contour plots by scanning gate microscopy \cite{SToussaint2018}. 

The arrangement in which we study the magnetoresistance structure at low $B$ is a confined TMF geometry. TMF is a quintessentially ballistic magnetotransport technique \cite{Tsoi1974Bi, Houten1989, Heremans1992, Tsoi1999, gupta2021NatComm, NanoResLett2022-17-Cole} that has been applied to the characterization of Fermi surfaces \cite{Tsoi1974Bi, Tsoi1975, Heremans1992, Tsoi1999, Taychatanapat, Lee2016, Berdyugin2020}, the measurement of electron-electron scattering \cite{gupta2021NatComm}, the detection of composite fermions \cite{Goldman1994, Smet1996}, and the study of Andreev reflection \cite{Tsoi1999} and spin-orbit interaction \cite{Heremans-AIP2007, Rendell-PRB2022}. 

\section{Device geometry and qualitative description}

\begin{figure}[!htbp]
\begin{center}
\includegraphics[width=15cm]{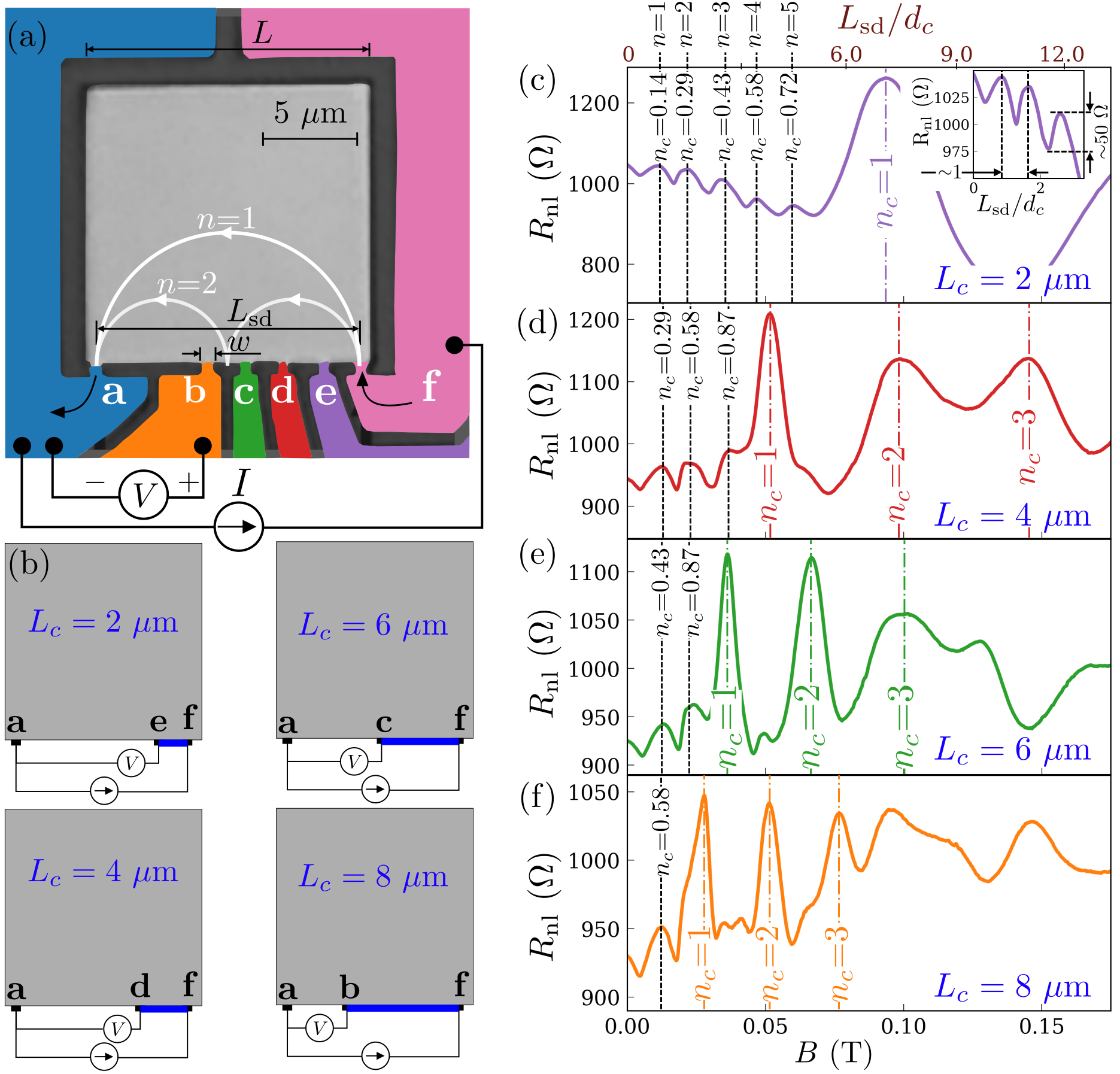}
\caption{(a) Optical micrograph of the device fabricated on the GaAs/AlGaAs heterostructure, with PCs a-f, where a = drain and f = source. Paths leading to each PC are colored differently, while black regions indicate etched barriers defining device boundaries. $L=15\,\mu \text{m}$, distance a-f = $L_\mathrm{sd}=13.8 \, \mu\text{m}$ and PC width $w=0.6 \, \mu\text{m}$. Current source ($I$, arrow indicates direction of conventional current) and voltage measurement ($V$) connections for $L_c$ = 8 $\mu$m are indicated. Semiclassical cyclotron orbits corresponding to $d_c = L_\mathrm{sd}/n$ where $n=1, 2$ are drawn in. (b) Measurement configurations for $L_c = 2, 4, 6, 8 \ \mu$m, with $L_c$ indicated by a blue trace at the bottom edge. (c-f) $R_\mathrm{nl}$ versus $B$ (bottom axis) and $L_\mathrm{sd}/d_c$ (top axis) at $T=$ 4.1 K, for $L_c = 2, 4, 6, 8 \ \mu$m (as in (b)). Inset in (c) shows $R_\mathrm{nl}$ versus $L_\mathrm{sd}/d_c$ zoomed into the fractional peaks. $\Delta (L_\mathrm{sd}/d_c) = \Delta n \approx 1$ and height of fractional peaks $\approx 50 \ \Omega$.}\label{fig:fig1}
\end{center}
\end{figure}

We consider a device geometry in a 2DES depicted in Fig.~\ref{fig:fig1}(a), a square defined by boundaries with internal edge length $L$, and with current source and drain point contacts (PCs) of conducting widths $w \ll L$ placed near two corners along the same edge at center-to-center distance $L_\mathrm{sd}$ (necessarily $L_\mathrm{sd}<L$). Conventional current is injected into the device through the source PC (corresponding to injection of holes, referring to positive charges due to empty states below the Fermi contour \cite{taubert-JAP2011, taubert-PRB2010, Firdaus2018, taubert2011}), and extracted through the drain PC (corresponding to injection of electrons, filling states above the Fermi contour). In the device under consideration (Fig.~\ref{fig:fig1}(a)) we have $L=15\,\mu \text{m}, \, L_\mathrm{sd}=13.8 \, \mu\text{m}$, and $w \approx 0.6\,\mu \text{m} = 0.04 \, L$. The scattering of electrons off the internal device boundaries is specular, as explained below. All measurements are performed at temperature $T=$ 4.1 K. In the presence of a perpendicular magnetic field $B$, carriers undergo semiclassical cyclotron orbits with cyclotron diameter $d_c =  2 \hbar k_F/(eB)$, where $e$ denotes the elementary charge, $\hbar$ the reduced Planck's constant, and $k_F$ the Fermi wavevector. The 3-terminal nonlocal resistance $R_\mathrm{nl}$ is defined as the voltage measured between a detector PC and the drain PC divided by the injected current $I$, where the detector PC is placed at a center-to-center distance $L_c < L_\mathrm{sd} < L$ from the source PC (Fig.~\ref{fig:fig1}(b)). $R_\mathrm{nl}$ reaches a positive peak whenever $d_c = L_c/n_c$, with $n_c$ an integer. These integer focusing peaks, observed in Fig.~\ref{fig:fig1}(c-f) for $n_c=1,2,3$, constitute TMF, and are a prototypical signature of ballistic transport. They can be understood in terms of single-particle ballistic trajectories of injected carriers which follow semiclassical cyclotron orbits and reach the detector PC after $n_c-1$ specular scattering events off the boundary between source and detector PC. Yet notably, the device geometry in Fig.~\ref{fig:fig1}(a) gives rise to an additional resonance condition visible as maxima in $R_\mathrm{nl}$: $d_c = L_\mathrm{sd}/n$, with $n$ an integer. This condition denotes a novel \emph{source-drain resonance}, corresponding to $d_c$ of a carrier trajectory injected at the source PC fitting an integer number of times into $L_\mathrm{sd}$ (Fig.~\ref{fig:fig1}(a)). The matching maxima in $R_\mathrm{nl}$ occur at values of $B$ independent of the location of the detector PC between source and drain. Figures~\ref{fig:fig1}(c-f) show that the device manifests the source-drain resonances experimentally at lower $B$, with distinct maxima in $R_\mathrm{nl}$ for $B$ that satisfy the resonance condition. The high-resolution simulations of ballistic transport likewise predict the resonances. The source-drain resonance peaks occur for all $L_c < L_\mathrm{sd}$, and therefore, at effectively \emph{fractional} $n_c = n(L_c/L_\mathrm{sd})$. As will be described, unlike the TMF origin of the integer $n_c$ peaks, the fractional source-drain resonance peaks cannot be attributed to any particular particle trajectory: \emph{they only occur from the collective dynamics arising from a particle distribution}. Further, while the integer peaks only require ballistic transport over the scale of $L_c$, the fractional peaks require more stringent device-scale ballistic transport along with specular boundaries. 

The experimental device in Fig.~\ref{fig:fig1}(a) is defined on a 2DES of very high electron $\mu$ hosted in a GaAs/AlGaAs heterostructure (Appendix A). Transport parameters of the 2DES are listed in Appendix A, including the long mean-free path $\ell = 64.5 \, \mu \textrm{m}$ at $T$ = 4.1 K. Hence $\ell >> L_\mathrm{sd}, L$, attesting to the very weak momentum-relaxing scattering which allows ballistic transport across the device. The Fermi wavelength $\lambda_F \simeq 43$ nm shows there are $N_m = w/(\lambda_F /2) \approx 28$ spin-degenerate transverse modes injected through the PCs into the device and hence transport through the PCs is effectively classical. Further, the wet-etching process used to define the boundaries in the 2DES (Appendix A) results in specular scattering \cite{gupta2021PRL, gupta2021NatComm, Heremans1999, Chen2005, Keser2021}, thus allowing the fractional peaks to manifest in the device. The 15 $\mu$m $\times$ 15 $\mu$m device in Fig.~\ref{fig:fig1}(a) features 6 PCs along the bottom boundary. Current is injected at the source PC at the right-end, and drained at the drain PC at the left-end. The intermediate PCs serve as detectors at $L_c = 2, 4, 6, 8$ $\mu$m (Fig.~\ref{fig:fig1}(b)). 

\section{Quantitative results and discussion}

Figures~\ref{fig:fig1}(c-f) show the nonlocal resistance $R_\mathrm{nl}$ measured at $T=4.1$ K as a function of $B$ and $L_\mathrm{sd}/d_c$. The integer $n_c=1,2,3$ peaks are conspicuous, e.g. for $L_c = 2$ $\mu$m the $n_c=1$ peak occurs at $B = 0.095$ T and for $L_c = 8$ $\mu$m the $n_c=3$ peak occurs at $B = 0.077$ T. Smaller clear peaks appear at lower $B$, below $n_c=1$, each with $R_\mathrm{nl}$ amplitude $\simeq 50$ $\Omega$ and at values of $d_c$ incommensurate with $L_c$. We note that firstly, more fractional peaks appear for a detector closer to the source, with detector PCs at $L_c = 2, 4, 6, 8$ $\mu$m detecting 5, 3, 2 and 1 fractional peaks respectively. This observation suggests that $L_c \ll d_c$ is a favorable condition for the fractional peaks to manifest. Secondly, the value of $B$ at which the fractional peaks occur is independent of $L_c$ (Fig.~\ref{fig:fig1}(c-f)). For example, we observe that the fractional $n=1$ peak occurs at the same $B$ for all detector PCs. In contrast, the integer $n_c = 1$ peak occurs at $B_{n_c=1} = (2\hbar k_F/e)(1/L_c)$ that depends on $L_c$. Finally, the periodicity of the fractional peaks is also independent of $L_c$, unlike the integer peaks the periodicity of which depends on $L_c$ as $\Delta B = (2\hbar k_F/e)(1/L_c)$ (Fig.~\ref{fig:fig1}(c-f)).

We compare the experimental results to simulations of ballistic magnetotransport based on the collisionless Boltzmann transport equation in the $T \rightarrow 0$ limit (Appendix B),
\begin{equation}
\hat{v}_F.\frac{\partial f}{\partial \mathbf{x}} + \left(\frac{2}{d_c}\right)\frac{\partial f}{\partial \theta}  = 0
\label{eq:boltzmann}
\end{equation}
where $f(\mathbf{x}, \theta)$ is the carrier probability distribution, $\mathbf{x} \equiv (x, y)$ are the spatial coordinates, $\theta$ is the angle on the (circular) Fermi contour and $\hat{v}_F \equiv (\cos(\theta), \sin(\theta))$ is the unit vector along the Fermi velocity. We solve Eq.~\ref{eq:boltzmann} using two numerical schemes: a finite volume (FV) scheme where the distribution is solved for on a uniform grid of the $(\mathbf{x}, \theta)$ coordinates, and a Monte-Carlo (MC) scheme where the trajectories of injected carriers are evolved by analytically computing their intersections with the device boundaries at each specular scattering event off the boundaries. The former is a ``bulk'' scheme, whose output is the distribution over the entire device, while the latter is a ``boundary'' scheme where the solution is confined to the device boundaries, and is therefore computationally faster than the bulk scheme. At the source PC, we impose $f(\theta) = \cos(\theta)/2$ where $\theta \in [-\pi/2, \pi/2]$ is the angle with respect to the normal to the device boundary, corresponding to injection of holes below the Fermi contour. Similarly, at the drain PC we impose $f(\theta) = -\cos(\theta)/2$, corresponding to injection of electrons above the Fermi contour \cite{chandra-arXiv2019, gupta2021PRL, gupta2021NatComm, taubert-JAP2011, taubert-PRB2010, Firdaus2018, taubert2011}. 

\begin{figure}[!htbp]
\begin{center}
\includegraphics[width=15cm]{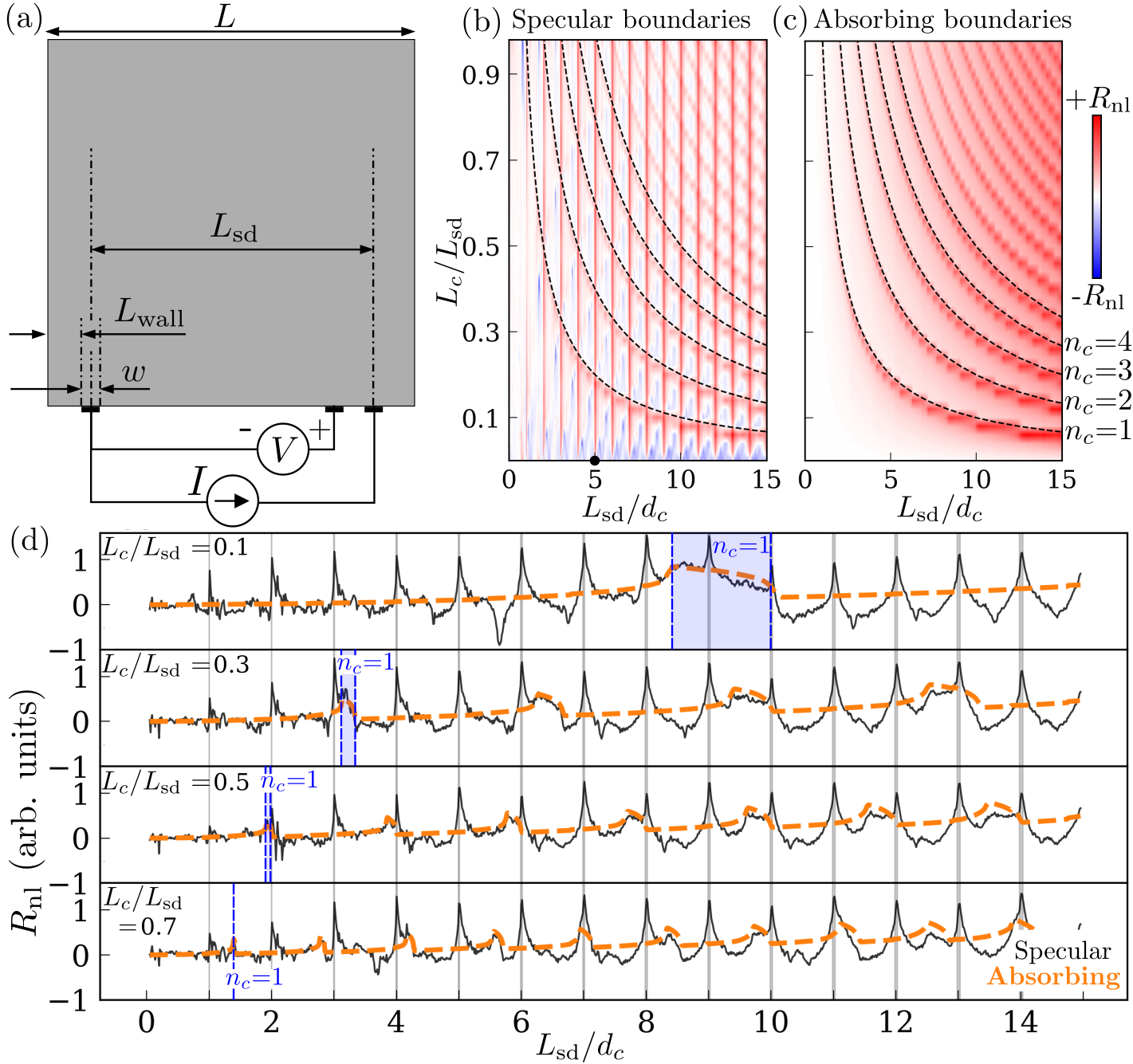}
\caption{(a) Device schematic for MC scheme simulations with $L$ (in arb. units), $w = 0.002L$ and $L_\mathrm{wall} = 0$. (b) Contour plot of simulated $R_{nl}$ vs $L_c/L_\mathrm{sd}$ and $L_\mathrm{sd}/d_c$ for specular device boundaries. $L_\mathrm{sd}$ is fixed while $d_c$ and $L_c$ are varied. (c) Contour plot as in (b) but with absorbing (non-specular) side and top boundaries, resulting in absence of fractional peaks. (d) $R_\mathrm{nl}$ vs $L_\mathrm{sd}/d_c$ for $L_c/L_\mathrm{sd} = 0.1, 0.3, 0.5, 0.7$ for specular boundaries (black, solid) and absorbing side and top boundaries (orange, dashed). In all panels blue vertical lines indicate the integer $n_c=1$ TMF peak (width indicated by shaded blue region bounded by vertical dashed lines). The width of the TMF peak is attributed to the finite width of source, drain and detector PCs ($w$), and decreases as $L_c$ is increased. Gray vertical bars indicate analytically expected locations of fractional peaks (at integer $L_{sd}/d_c$) and their thickness signifies the spread in the expected locations due to finite $w$. Unlike TMF peaks, fractional peaks require specular boundaries.}\label{fig:fig2}.
\end{center}
\end{figure}

We perform simulations in a square domain of side $L$ (in arbitrary units), with varying PC widths $w = 0.002L - 0.02L$ (Appendix C), and for varying source and drain PC distances from the sidewalls (Appendix D), described by $L_\mathrm{wall} = \frac{1}{2} (L-L_\mathrm{sd} - w)$ (device schematic in Fig.~\ref{fig:fig2}(a)). The simulations also consider specular as well as absorbing (non-specular) boundaries (Fig.~\ref{fig:fig2}). In the actual experiment's device as mentioned we have $L=15\,\mu \text{m}, \, L_\mathrm{sd}=13.8 \, \mu\text{m}, \, w = 0.04 \, L = 0.6 \, \mu\text{m}$ and hence $L_\mathrm{wall} = 0.3 \, \mu\text{m}$, and its boundaries are specular. In the simulations we consider the fiducial case of $w = 0.002L$ for all three PCs and $L_\mathrm{wall} = 0$, corresponding to source and drain PCs flush with the sidewalls and $L_\mathrm{sd}=L-w$. Figure~\ref{fig:fig2}(b) shows a contour plot, simulated in the MC scheme, of $R_\mathrm{nl}(L_c/L_\mathrm{sd}, L_\mathrm{sd}/d_c)$ where $L_\mathrm{sd}$ is constant while $d_c$ and $L_c$ are varied, for a device with specular boundaries, measured using a fictitious detector PC centered at $L_c$. The intensity of red (blue) color in the contour plots indicate the magnitude of positive (negative) $R_\mathrm{nl}$ measured by the detector. The locations of the center of the source and drain PCs correspond to $L_c/L_\mathrm{sd}=0$ and $L_c/L_\mathrm{sd}=1$ (for which $L_c=L-w$ if $L_\mathrm{wall} = 0$) on the vertical axis, respectively. The integer $n_c$ TMF peaks appear as hyperbolas obeying $(L_c/L_\mathrm{sd})(L_\mathrm{sd}/d_c)=L_c/d_c=n_c$. The fractional peaks at integer $n$ appear as vertical features, independent of $L_c/L_\mathrm{sd}$ at every $L_\mathrm{sd}/d_c = n$ (see e. g. $n = 5$ indicated by a black dot in Fig.~\ref{fig:fig2}(b)). When $R_\mathrm{nl}$ is computed for varying $L_\mathrm{sd}/d_c \propto B$ at a fixed $L_c/L_\mathrm{sd}$, corresponding to a horizontal cut in the contour plot, the source-drain resonances appear as peaks at fractional $n_c = n(L_c/L_\mathrm{sd})$. These peaks are depicted vs $L_\mathrm{sd}/d_c \propto B$ in Fig.~\ref{fig:fig2}(d), for better visualization of Fig.~\ref{fig:fig2}(b). By considering Figs.~\ref{fig:fig2}(a,b,d), we are able to explain features observed in the experimental data. First, the intersections of the vertical lines (due to source-drain resonances) with the hyperbolae of integer TMF peaks reveals why detector PCs closer to the source detect more fractional peaks. For a detector PC with $L_c/L_\mathrm{sd} = 0.1$, the fractional peaks occur at $n_c = 0.1n$, yielding 9 fractional peaks below $n_c = 1$ (including the one at $n = 9$, which lies within the shaded blue region in Fig.~\ref{fig:fig2}(d)). For a PC at $L_c/L_\mathrm{sd} = 0.3$, there are only 3 such peaks. Second, the values of $L_\mathrm{sd}/d_c \propto B$ at which the fractional peaks occur are indeed independent of $L_c$, depending instead on the device-scale $L_\mathrm{sd}$. As a result, the periodicity of the fractional peaks is also independent of $L_c$, unlike the integer TMF peaks whose periodicity is $\propto 1/L_c$. In Fig.~\ref{fig:fig2}(c), we consider the same device, but with the side and top boundaries absorbing rather than specular. The fractional peaks are now completely absent, with only the integer $n_c$ TMF peaks visible in $R_\mathrm{nl}(L_c/L_\mathrm{sd}, L_\mathrm{sd}/d_c)$. The absence of the fractional peaks signifies that device-scale specular boundary scattering is also necessary, unlike for integer TMF peaks which only require specular scattering off the bottom boundary. 

In Fig.~\ref{fig:fig2}(b) where $w = 0.002L$, the modeled oscillation amplitude of $R_\mathrm{nl}$ for the fractional peaks is larger than of the integer peaks, unlike the experimentally measured amplitude in Fig.~\ref{fig:fig1}(c-f). To explore the change, $R_\mathrm{nl}$ was modeled in the MC scheme for variable source and drain PC width $w$ in Appendix C. The model indicates that as $w$ is increased, the fractional peaks lose prominence and eventually the amplitudes of the integer peaks exceed those of the fractional peaks. The experimental findings in Fig.~\ref{fig:fig1}(c-f) hence are explained by relatively wider $w$. The effects of larger source and drain PC distances $L_\mathrm{wall}$ from the sidewalls were also modeled in Appendix D. Increasing $L_\mathrm{wall}$ substantially diffuses the fractional peaks, contributing to the observations in Fig.~\ref{fig:fig1}(c-f). Hence narrow $w$ and small $L_\mathrm{wall}$ are beneficial for the fractional peaks to manifest. We have thus achieved a minimal model that reproduces the essential features of the experiment. 

We performed a quantitative correlation between the experimental data and the model, via comparison of $B$-values at which the fractional peaks occur in the experiment and the model. The model predicts a periodicity in $B$ given by $\Delta (L_\mathrm{sd}/d_c) = \Delta n = 1$, where $L_\mathrm{sd}/d_c \propto B$. Analysis of the periodicity in $B$ in the data in Fig.~\ref{fig:fig1}(c-f) allows calculation of the device's $L_\mathrm{sd}$, in which any small systematic offsets in the experimental $B$ cancel out. This value of $L_\mathrm{sd}$ can be vetted against the lithographic length. Analysis of the data in Fig.~\ref{fig:fig1}(c-f) averaging over all PC combinations, yields $L_\mathrm{sd} = 15.8 \pm 3.1 \ \mu $m, in good agreement with the lithographic length $L_\mathrm{sd} = 13.8 \ \mu$m in Fig.~\ref{fig:fig1}(a).

\begin{figure}[!htbp]
\begin{center}
\includegraphics[width=16cm]{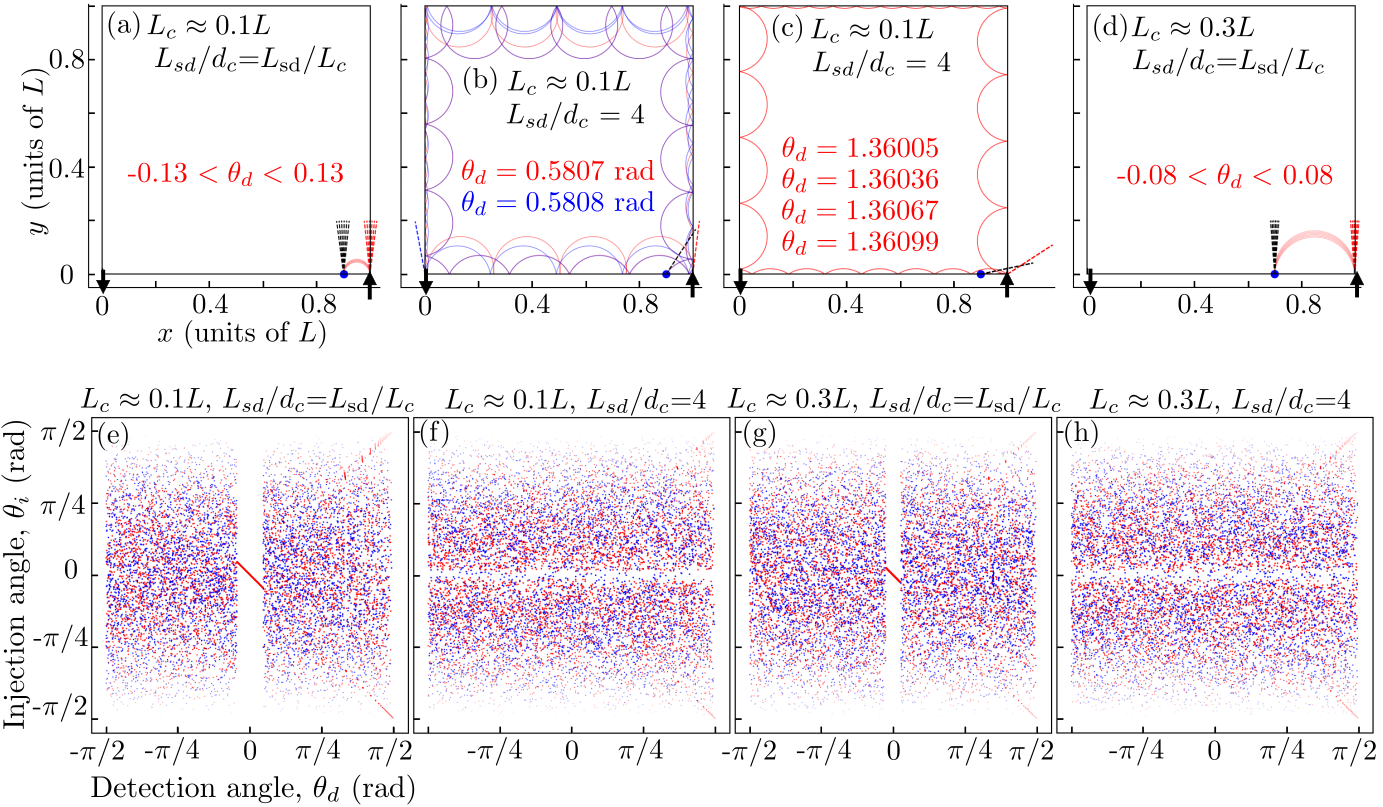}
\caption{(a) Trajectories corresponding to the $n_c = 1$ TMF peak in a square device of size $L \times L$, simulated using the MC scheme. Cyclotron orbits from the source PC (at right) land directly on the detector PC (indicated by a blue dot), here at $L_c\approx 0.1 L$. Red (black) dashed lines show tangents to the trajectories at the source (detector), and $\theta_d$ denotes the arrival angle at the detector. (b) Two trajectories arriving at the detector PC at close angles $\theta_d$ can have traced very different paths through the device. The red (blue) trajectory indicates a hole (electron) injected from the source (drain, at left). (c) Trajectories (red) arriving at the detector PC over a range of $\theta_d$ can also arise from non-diverging nearly identical paths through the device. (d) Trajectories corresponding to the $n_c = 1$ TMF peak with the detector location at $L_c \approx 0.3 L$. Trajectories injected within a smaller $\theta_i$ span now reach the detector. (e) ARTS of $\theta_i$ (injection angle at source PC (hole trajectories, red) or drain PC (electron trajectories, blue)) versus arrival angle $\theta_d$ at detector PC, for the $n_c = 1$ TMF peak for $L_c \approx 0.1L$, as in (a). Contiguous-dot structure at $\theta_d,\theta_i \approx 0$ corresponds to the $n_c = 1$ TMF peak. (f) ARTS for the fractional $n = 4$ ($n_c = 0.397$) peak for $L_c \approx 0.1 L$ as in (b,c). The gap at $\theta_i \approx 0$ correlates with the appearance of the fractional peak. No noticeable contiguous-dot patterns are present. (g) ARTS for the $n_c = 1$ TMF peak for $L_c \approx 0.3L$, as in (d), showing that the contiguous-dot structure corresponding to the $n_c = 1$ TMF peak lies within a tighter span in $\theta_i$ compared to (e). (h) ARTS for the fractional $n = 4$ ($n_c = 0.397$) peak for $L_c \approx 0.3L$.} \label{fig:fig3}
\end{center}
\end{figure}

Individual single-particle trajectories can shed light on the unique origins of the fractional peaks, as illustrated in Fig.~\ref{fig:fig3} where a square device of internal dimensions $L$ is modeled in the MC scheme. Figure~\ref{fig:fig3} examines the trajectories that land in the detector PC, due to carriers originating from either the source (holes) or from the drain (electrons), using an approach labeled angle-resolved trajectory spectroscopy (ARTS, details in Appendix B). In Figs.~\ref{fig:fig3}(a,b,c,d) all PCs have $w = 0.002L$, $L_\mathrm{wall} = 0$, while the source PC is centered at $0.999L$, and the drain PC centered at $0.001L$ ($L_\mathrm{sd} = L-w = 0.998 \, L \approx L$, source and drain PCs flush with the sidewalls). The detector PC is centered at $0.9L$ (hence $L_c = 0.099 \, L \approx 0.1 \, L$) for Fig.~\ref{fig:fig3}(a,b,c,e,f), while it is centered at $0.7L$ ($L_c = 0.299 \, L \approx 0.3 \, L$) for Fig.~\ref{fig:fig3}(d,g,h). In Fig.~\ref{fig:fig3}(a), $d_c = L_c = 0.099 \, L$ is chosen to correspond to the condition for observation of the integer $n_c=1$ TMF peak. Shown are a set of trajectories emanating from the source, with small injection angles $\theta_i$ centered around zero (tangents indicated by dashed red lines). These trajectories correspond to the usual single-particle intuition associated with the integer TMF peaks, with the spread in the injection angles due to the finite sizes of the source and detector PCs and the angular spread of trajectories due to the Fermi contour. Such trajectories are relatively insensitive to $\theta_i \approx 0$, robustly landing at the detector PC after following a direct cyclotron-orbit path from source PC to detector PC, conforming to TMF. The dashed black lines indicate tangents to the trajectories arriving over arrival angles $\theta_d$ at the detector PC, for $-0.13 < \theta_d < 0.13$ (in rad). In Figs.~\ref{fig:fig3}(b-c), $d_c = L_\mathrm{sd}/4 = 0.2495 \, L$ is chosen to correspond to the condition for observation of the fractional $n=4$ ($n_c = 0.397$) peak. Figure~\ref{fig:fig3}(b) illustrates two trajectories arriving at the detector PC at very nearly identical $\theta_d \approx 0.5807...0.5808$ (tangent indicated by dashed black line), but having traversed very different trajectories through the device: the red trajectory follows a hole injected from the source PC (at $\theta_i$ indicated by dashed red line), the blue trajectory follows an electron injected from the drain PC (at different $\theta_i$ indicated by dashed blue line). Yet Fig.~\ref{fig:fig3}(c) reveals that also trajectories exist which arrive at the detector PC over a small range of $\theta_d$ (tangents indicated by single dashed black line and values of $\theta_d$ indicated in the figure), but originate from still closely bunched non-diverging trajectories (red trajectories following holes injected from the source PC, $\theta_i$ values indicated by single dashed red line). Figure~\ref{fig:fig3}(d) shows trajectories at $\theta_i \approx 0$ corresponding to the integer $n_c = 1$ TMF peak, but now for a different detector location, $L_c \approx 0.3 L$. The figure shows that the span of $\theta_i$ for which injected trajectories reach the detector decreases as the detector PC is moved further away from the source PC. 

The ARTS plots in Figs.~\ref{fig:fig3}(e-h) show the injection angle $\theta_i$ at source (hole trajectories, coded in red) or drain (electron trajectories, coded in blue) PCs vs the arrival angle $\theta_d$ at the detector PC. The opacity of the dots is weighted by $\cos(\theta_i)$, reflecting the injection distribution over $\theta_i$ in the MC scheme (fewer trajectories injected at high $|\theta_i|$ also mean fewer detected at the detector PC). Figure~\ref{fig:fig3}(e) corresponds to Fig.~\ref{fig:fig3}(a), with $d_c = L_c$ conforming to the condition for observing the integer $n_c=1$ TMF peak, while Fig.~\ref{fig:fig3}(f) corresponds to Figs.~\ref{fig:fig3}(b-c), with $d_c = L_\mathrm{sd}/4$ conforming to the condition for observing the fractional $n=4$ ($n_c = 0.397$) peak. In Fig.~\ref{fig:fig3}(e) the contiguous-dot structure observed for $\theta_d, \theta_i \approx 0$ indicates the integer $n_c = 1$ TMF peak, and is attributed to the trajectories depicted in Fig.~\ref{fig:fig3}(a): over a small range of $\theta_i$ trajectories reliably land at the detector PC in a one-to-one relation to $\theta_d$. The width of the contiguous-dot structure is proportional to $w$. Less apparent contiguous-dot structures also exist in Fig.~\ref{fig:fig3}(e) for $\theta_d \lessgtr 0$, corresponding to trajectories originating from the source or drain PCs and reaching the detector PC with more than one specular scattering event off the boundaries. Notably, between the contiguous regions in Fig.~\ref{fig:fig3}(e) exists a highly unordered pointillistic structure with a mix of electron and hole trajectories. The spatial profiles of trajectories in the unordered regions reveal that such trajectories that land on the detector PC at infinitesimally close $\theta_d$ can have uncorrelated $\theta_i$, reflecting similar observations in Fig.~\ref{fig:fig3}(b). The contiguous-dot structure at $\theta_d, \theta_i \approx 0$ originating in the integer $n_c = 1$ TMF peak provides the dominant contribution to $R_\mathrm{nl}$ at the detector PC. The unordered regions do not cooperate to contribute a sizable signal. In Fig.~\ref{fig:fig3}(f) the horizontal blurred gap over all $\theta_d$ for $\theta_i \approx 0$ correlates with the fractional $n=4$ ($n_c = 0.397$) peak (explored for other $n$ in Appendix F). Example trajectories are shown in Figs.~\ref{fig:fig3}(b-c). The ARTS structure remains overall unordered, without noticeable contiguous-dot patterns, indicating mostly divergent trajectories. Hence, apart from the gap, strikingly ARTS reveals an \emph{absence} of any contiguous structure to which the fractional peak can be attributed, unlike in the case of the integer TMF peak. The gap indicates a dearth of trajectories reaching the detector PC. This dearth is more acutely sensed in transport because the number of injected trajectories is weighted by $\cos(\theta_i)$, maximal for $\theta_i \approx 0$, congruent with the appearance of a fractional peak in measurements and modeling.  We next consider a different detector location, $L_c \approx 0.3 L$. In Fig.~\ref{fig:fig3}(g), we set $d_c = L_c$, conforming to the condition for observing the integer $n_c = 1$ TMF peak. We again observe the contiguous-dot structure arising from trajectories near $\theta_i \approx 0$ landing directly at the detector PC as expected at the integer $n_c = 1$ TMF peak. Comparison with Fig.~\ref{fig:fig3}(e) shows that the width of the contiguous-dot structure is inversely related to $L_c$, meaning that trajectories within a tighter span of $\theta_i$ reach the detector, consistent with Fig.~\ref{fig:fig3}(d). In Fig.~\ref{fig:fig3}(h) we consider the condition for the $n=4$ fractional peak again, but for $L_c \approx 0.3$. The blurred gap near $\theta_i \approx 0$ is again observed, indicating a dearth of trajectories reaching the detector PC. The ARTS structure overall appears strikingly similar to Fig.~\ref{fig:fig3}(f), hinting that the phenomenology of fractional peaks is independent of the detector PC location. We infer that the source-drain resonance results in an increased number of hole (electron) trajectories finding an expeditious path from source (drain) PC to drain (source) PC, rapidly exiting the device and never arriving at the detector PC irrespective of its location. While the ARTS gap is striking, the exact mechanism of its relation to the formation of maxima in the nonlocal $R_\mathrm{nl}$ at the detector PC bears investigating in future work. Even the trajectories corresponding to the fewer dots within the blurred gap trace a complex trajectory through the device. The lack of dominant contribution from an ordered contiguous-dot structure (such as for TMF) precludes an intuitive explanation based on obvious single-particle trajectories. The fractional peaks in $R_\mathrm{nl}$ appear to build from many random events in the collective dynamics of the particle system. Fractional peaks hence require consideration of \emph{all} trajectories and the collective dynamics of a particle distribution which includes electrons and holes. In contrast, the integer TMF peaks can be understood from more intuitive single-particle semiclassical ballistic cyclotron orbits with $\theta_d, \theta_i \approx 0$.

\section{Correlation with backflow and vorticity}

\begin{figure}[!htbp]
\begin{center} 
\includegraphics[width=11cm]{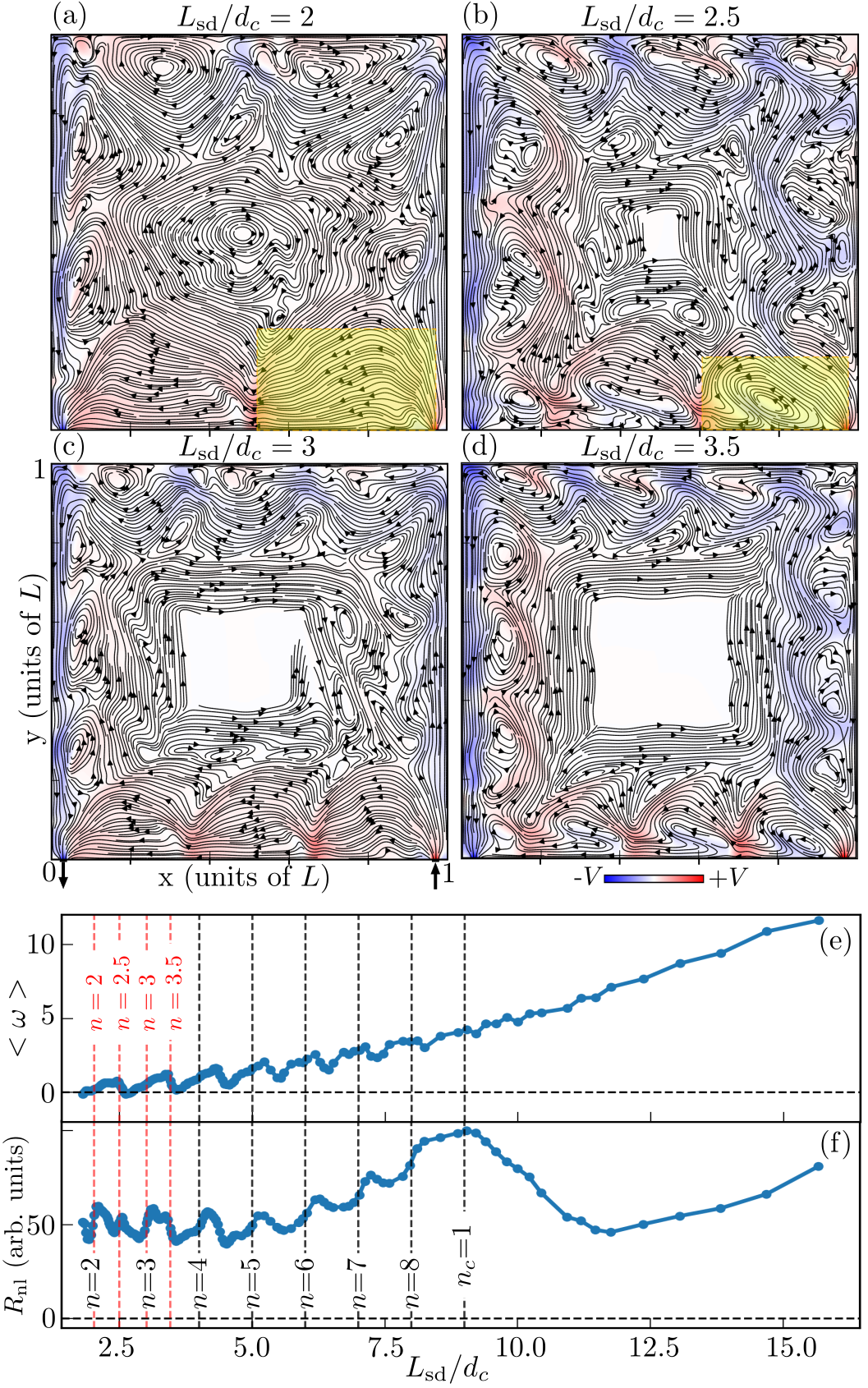}
\caption{Current streamlines and voltage contour plots for an $L \times L$ device with specular boundaries, with $L_\mathrm{sd} = 0.94L$, $L_\mathrm{wall} = 0.02L$ and $w = 0.02L$ for (a) $L_\mathrm{sd}/d_c = 2$, (b) $L_\mathrm{sd}/d_c = 2.5$, (c) $L_\mathrm{sd}/d_c = 3$, and (d) $L_\mathrm{sd}/d_c = 3.5$. (e) Averaged vorticity $\langle \omega \rangle$ vs $L_\mathrm{sd}/d_c$ (averaging regions $d_c \times d_c/2$ yellow shaded in (a) and (b)). Vertical dashed black lines indicate integer $L_\mathrm{sd}/d_c$ (expected location of fractional peaks in $R_\mathrm{nl}$), while red lines indicate positions of (a-d) along $L_\mathrm{sd}/d_c = n$. (f) $R_\mathrm{nl}$ vs $L_\mathrm{sd}/d_c$ for $L_c = 0.102 L$, for comparison with $\langle \omega \rangle$. Fractional peaks in $R_\mathrm{nl}$ occur near integer $L_\mathrm{sd}/d_c$, with systematic shifts due to finite $w$ (Appendix C). The periodicity in $\langle \omega \rangle$ nearly equals that in $R_\mathrm{nl}$. $n_c = 1$ corresponds to the first TMF peak.}\label{fig:fig4}
\end{center}
\end{figure}

Another striking manifestation of the source-drain resonance condition is contained in Fig.~\ref{fig:fig4}, where we examine the relation between the spatial transport profile and the source-drain resonances. Figures~\ref{fig:fig4}(a-d) are obtained via the FV scheme and show current streamlines and color-coded voltage contour plots in the steady-state for an $L \times L$ device with specular boundaries, with $L_\mathrm{sd} = 0.94L$, $L_\mathrm{wall} = 0.02L$ and $w = 0.02L$, at $L_\mathrm{sd}/d_c$ = 2, 2.5, 3 and 3.5 respectively. As mentioned previously, the ballistic regime can exhibit current vortices, characterized by a non-zero vorticity $\omega = \nabla \times \mathbf{j}$, where $\mathbf{j}$ denotes the current density. The vector sum of many individual electron and hole trajectories over time gives rise to $\mathbf{j}$ in the steady-state, which is hence the outcome of collective dynamics more intricate than understood from single-particle cyclotron orbits. Figure~\ref{fig:fig4}(a) examines the profiles for the fractional $n=2$ peak, at $L_\mathrm{sd}/d_c=2$. A focusing of current streamlines occurs at the center of the bottom edge, with the conventional current flow from the source indicated by arrows. In keeping with the previously used color scheme, a red shading indicates a local overdensity of holes and a positive voltage, while blue shading indicates a local overdensity of electrons and a negative voltage. Fig.~\ref{fig:fig4}(a) shows an overdensity of holes and a positive voltage everywhere throughout the bottom edge. This overdensity of holes and positive voltage correspond to the fractional $n=2$ peak, as expected since fractional peaks occur independent of detector location $L_c$. The associated current flow is unidirectional from source to the drain along the bottom edge. Figure~\ref{fig:fig4}(b) examines the profiles at $n=L_\mathrm{sd}/d_c=2.5$. The focusing of current streamlines occurs where expected at the bottom edge but the overdensity of holes is around the center of the semiclassical cyclotron orbits (e.g. between the source and the first focusing location) now replaced with an overdensity of electrons and a locally negative voltage. The electrons originate from the drain and traverse the left, top and right boundaries. The overdensity of electrons and negative voltage result in a local backflow of the current, resulting in a striking phenomenon: the appearance of current vortices. Figures~\ref{fig:fig4}(c,d) repeat and confirm the phenomena, for the fractional $n=L_\mathrm{sd}/d_c=3$ peak and for the half-integer $n=L_\mathrm{sd}/d_c=3.5$ condition. At integer $n=L_\mathrm{sd}/d_c$, corresponding to a fractional peak, the voltage along the bottom edge does not alternate sign (Figs.~\ref{fig:fig4}(a,c)), while at half-integer $n=L_\mathrm{sd}/d_c$ the voltage along the bottom edge alternates sign, causing backflow and current vortices (Figs.~\ref{fig:fig4}(b,d)). 

Figures~\ref{fig:fig4}(e,f) are also obtained via the FV scheme. Figure~\ref{fig:fig4}(e) considers the vorticity averaged over an area $d_c \times d_c$/2, corresponding to a box covering one cyclotron orbit with its right-bottom corner located at the midpoint of the source PC (indicated by shaded yellow region in Figs.~\ref{fig:fig4}(a,b)), and shows the area-averaged vorticity $\langle \omega \rangle = \int \omega \, dx \, dy$ vs $L_\mathrm{sd}/d_c$. Remarkably, Fig.~\ref{fig:fig4}(e) shows that periodic features appear in $\langle \omega \rangle$ vs $L_\mathrm{sd}/d_c$ and that $\langle \omega \rangle$ peaks for half-integer $L_\mathrm{sd}/d_c$. The spatial profiles of $\mathbf{j}$ in also reveal the presence of current vortices along the bottom edge around half-integer values of $L_\mathrm{sd}/d_c$ (Figs.~\ref{fig:fig4}(b,d)) and the absence of vortices at integer values of $L_\mathrm{sd}/d_c$ (Figs.~\ref{fig:fig4}(a,c)). As noted, the absence of $\langle \omega \rangle$ at integer $L_\mathrm{sd}/d_c$ corresponds to the existence of positive voltages throughout the bottom edge, in turn corresponding to the fractional peak in $R_\mathrm{nl}$ independent of $L_c$. At half-integer $L_\mathrm{sd}/d_c$ locally negative voltages along the bottom edge cause the current backflow required for $\langle \omega \rangle$. In Fig.~\ref{fig:fig4}(f), $R_\mathrm{nl}$ is plotted vs $L_\mathrm{sd}/d_c$ for an arbitrarily chosen $L_c = 0.102 L$. In Fig.~\ref{fig:fig4}(e) the periodicity in $\langle \omega \rangle$ is found to be $\Delta (L_\mathrm{sd}/d_c)_{\langle \omega \rangle} = 1.077 \pm 0.121 \approx \Delta n = 1$, equalling the periodicity expected for the fractional peaks in $R_\mathrm{nl}$. This indicates a strong correlation between $\langle \omega \rangle$ (a property of the bulk of the device, not directly visualized experimentally) and $R_{nl}$ (a property measured at the device boundary). At present a complete physical insight for the correlation between $\langle \omega \rangle$ and $R_\mathrm{nl}$ is beyond the scope of this work. However the correlation underlines the importance of collective dynamics in ballistic transport, and highlights a complexity reaching beyond single-particle trajectories.  

\section{Implications and conclusions}

The experimental measurements and simulations both point to the existence of hitherto unsuspected structure in the nonlocal ballistic magnetoresistance $R_\mathrm{nl} (B)$ in a square mesoscopic geometry with source, drain and voltage detection PCs, similar to the TMF measurement geometry. $R_\mathrm{nl} (B)$ vs $L_\mathrm{sd}/d_c \propto B$, recorded at the detector PC, shows a periodic structure at low $B$, with peaks at locations where $d_c$ obeys $L_\mathrm{sd}/d_c = n$, where $n = 1,2,3..$. Since $L_\mathrm{sd} \neq L_c$, the peaks occur at fractional $n_c = L_c/d_c$, unlike the TMF peaks that occur at integer $n_c$. The fractional peaks occur at source-drain resonances where in a semiclassical cyclotron orbit framework, injected carriers from the source PC would impinge directly on the drain PC after $n-1$ specular reflections off the device boundary. However unlike TMF peaks, fractional peaks cannot be explained by single-particle ballistic trajectories. The particle trajectories leading to the fractional peaks are complex, and no single dominant contribution from a coordinated ordered group of trajectories is identifiable. The lack of dominant coordinated trajectory contribution forestalls an explanation based on conspicuous single-particle trajectories. Instead, the fractional peaks in $R_\mathrm{nl} (B)$ vs $B$ originate in the collective dynamics of the fermionic particle distribution, including electrons and holes, and require analysis and averaging of all trajectories. 

Further the source-drain resonances and fractional peaks (boundary properties) are anticorrelated with vorticity in the current density streamlines, a bulk property and quintessentially a collective flow phenomenon. The current density itself results from the time-averaged collective sum of many individual electron and hole trajectories. The amplitude of the fractional peaks is sensitive to the locations of source and drain PCs and to PC conducting widths, but not to the location of the detector PC. The insensitivity to the location of the detector PC is related to the disappearance of vorticity along the bottom boundary at the condition of the fractional peaks. 

The notable match between experiments and simulations supports the physical understanding that has been gained. The simulations can hence lead to new experimentally verifiable phenomena in ballistic transport in 2DESs and 2D hole systems or in recent materials with long carrier mean-free paths. The combined experiments and simulations can in future work also shed light on less understood and experimentally less-accessible aspects of low-dimensional transport, particularly vorticity. A main outcome of the work lies in the realization that ballistic transport results from a collective dynamics of a particle distribution and supports phenomena substantially more elaborate than those ostensibly deduced from single-particle cyclotron orbits.

\section*{Acknowledgements}
The authors acknowledge computational resources (GPU cluster \texttt{infer}) and technical support provided by Advanced Research Computing at Virginia Tech and the Center for Computational Innovations at Rensselaer Polytechnic Institute. MJM acknowledges support from the U.S. Department of Energy, Office of Science, Office of Basic Energy Sciences, under award number DE-SC0020138. MC and RS acknowledge support from the National Science Foundation under grant No. DMR-1956015.

\section{Appendices}

\subsection{Appendix A: Experimental methods}

The device is fabricated from a high-$\mu$ GaAs/AlGaAs heterostructure epitaxially grown using optimized molecular beam epitaxy. The 2DES resides in the GaAs quantum well, with depth (from the surface) of 190 nm and width 26 nm. The heterostructure uses a stepped barrier with Al alloy fractions 23\% and 32\% and is top- (9 $\times$ 10$^{11}$ cm$^{-2}$) and bottom-doped (4 $\times$ 10$^{11}$ cm$^{-2}$) using Si $\delta-$layers in Al$_{0.32}$Ga$_{0.68}$As with a setback of 80 nm. The growth and optimization of similar heterostructures is discussed in Ref. \cite{Gardner2016}. Samples in the van der Pauw configuration provide the transport parameters of the material under the same conditions as the device measurements, as listed in Table~\ref{tab:trans}. The transport parameters are derived from sheet resistance and Hall measurements accounting for conduction band non-parabolicity with a $\Gamma$-point effective mass $m=0.067 \, m_e$ where $m_e$ denotes the free electron mass. In particular, $\ell = 64.5 \, \mu$m is much larger than the device size $\sim 15 \, \mu$m, and hence transport through the device is in the ballistic regime. 

\begin{table}[!htbp]
\centering
	\caption{Transport parameters at $T$ = 4.1 K}
	\label{tab:trans}
\begin{tabular}{ |p{4.2cm}|p{2.6cm}|}
 \hline
Sheet resistance & 2.74 $\Omega / \square$ \\
2D electron density $N_S$ & 3.40$\times 10^{15}$ m$^{-2}$ \\
Mobility $\mu$ & 670 m$^2$/Vs  \\
Fermi wavevector $k_F$ & 1.46$\times 10^{8}$ m$^{-1}$ \\
Fermi wavelength $\lambda_F$ & 43.0 nm \\
Fermi velocity $v_F$ & 2.46$\times 10^{5}$ m/s \\
Fermi energy $E_F$ & 11.9 meV \\
Mean-free path $\ell$ & 64.5 $\mu$m \\
 \hline
\end{tabular}
\end{table}

For device fabrication, a Hall mesa is first patterned using photolithography followed by wet etching in H$_2$SO$_4$/H$_2$O$_2$/H$_2$O solution. The active region of the sample containing the square mesoscopic device is patterned using electron beam lithography using PMMA as the etching mask, followed by gentle and shallower wet etching in the same solution, resulting in smooth device boundaries. The depletion layer ($\sim$200 nm) forming between 2DES and etched boundaries smoothens out the potential over length scales smaller than the Fermi wavelength $\lambda_F$ by exponentially attenuating the high spatial frequency components of the potential. This results in specular boundaries. N-type Ohmic contacts are annealed InSn. The device is measured in a cryostat at $T$ = 4.1 K after LED illumination and stabilization. All measurements are obtained using low-frequency lock-in techniques under low ac current excitation (100 nA to 200 nA). To experimentally detect and resolve the fractional  peaks, very precise measurements of $B$ are needed, achieved using a gaussmeter and high-resolution stepping of the magnet current in each experimental run. 

\subsection{Appendix B: Transport simulations}
\
To simulate magnetotransport in the devices in the FV scheme \texttt{BOLT} \cite{chandra2019} is used, a solver framework for the Boltzmann transport equation (Eq.~\ref{eq:boltzmann}). The modeling takes the band structure of the material and device geometry (spatial and momentum space shape) as input. A square device is considered, with spatial dimensions $L \times L$, discretized into 250 $\times$ 250 numerical zones. In momentum space, we take the zero-temperature limit and hence the carriers are confined to the nearly circular Fermi contour, which is discretized using 1024 numerical zones. This 1D momentum space allows the simulations to be sped up significantly. The model assumes that transport is collisionless (i.e., there are no momentum-relaxing or momentum-conserving electron-electron interactions), and the system is in the ideal ballistic regime. The device boundaries are assumed to be perfectly specular. Current injection (extraction) is achieved by imposing a shifted Fermi-Dirac distribution at the location of the source (drain) contacts. The device is initialized with a thermal distribution everywhere else, which is then evolved as a function of time until steady state is reached. Currents and voltages at any time instant can then be calculated from the carrier distribution function as described in \cite{chandra2019}. 

In the MC scheme, holes (electrons) are injected through the source (drain) by randomly sampling the distribution $P(\theta_i) = 0.5\cos(\theta_i)$ with $N = 2\times 10^5$ particles. The trajectory of each injected particle is evolved analytically by computing the intersections of the circular orbits with the device boundaries, until the particle exits through either the source or the drain. For the ARTS (angle-resolved trajectory spectroscopy) calculations, we discretize the arrival angle, $\theta_d$, into $N=100000$ equally-spaced angles, each of which generates a single trajectory. We then analytically integrate this trajectory (as in the MC scheme) backwards till it reaches the source or the drain. We note the final angle $\theta_i$ which the trajectory makes upon striking the source or drain, which we call the injection angle. Each point in the ARTS plots (Figs.~\ref{fig:fig3}(d,e)) corresponds to a single trajectory measured by the detector, with its size weighted by a $\cos(\theta_i)$ factor. The uniform discretization in the ARTS scheme is essential for a key point: the random distribution of trajectories in the ARTS plots is not due to a random sampling of the trajectories, but rather due to a deterministic outcome obtained by analytically evolving trajectories that are equally spaced on the Fermi contour. 

\begin{figure}[!htbp]
\begin{center}
\includegraphics[width=15cm]{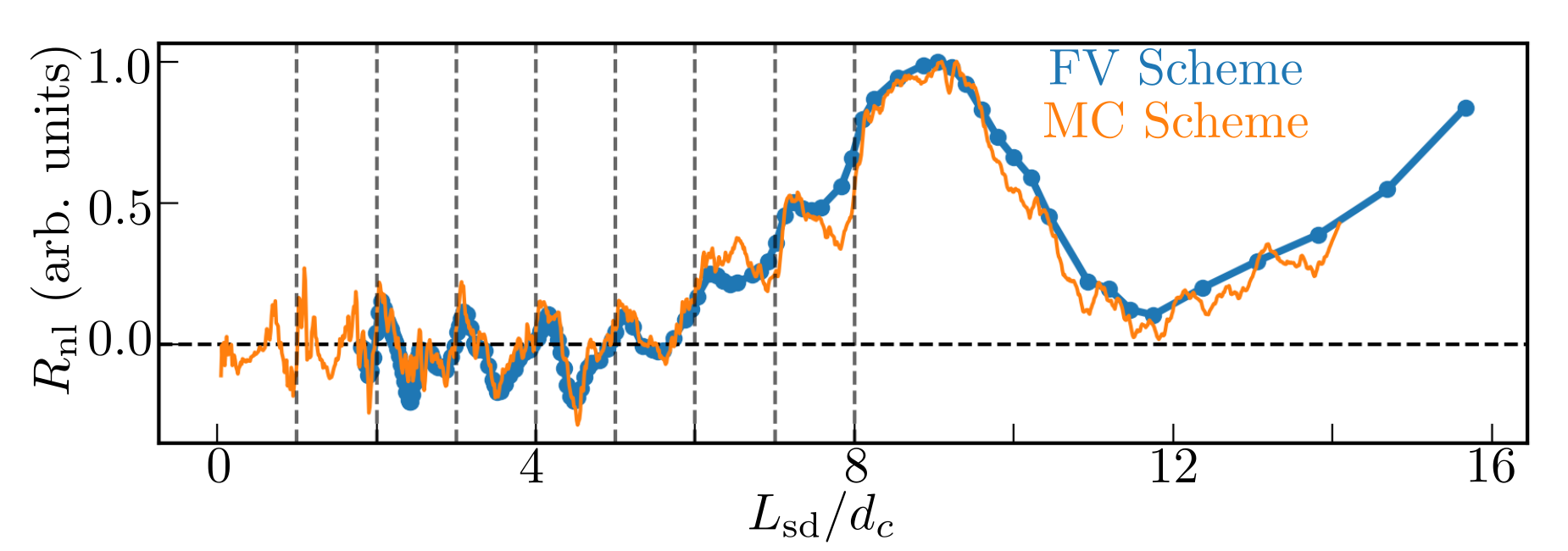}
\caption{Comparison of normalized simulated $R_\mathrm{nl}$ (in arbitrary units) vs $L_\mathrm{sd}/d_c$ obtained via the FV scheme (blue) and the MC scheme (orange) at $L_c = 0.1L$, in a $L \times L$ device with $w = 0.02L$ , $L_\mathrm{wall} = 0.02L$ and $L_\mathrm{sd} = 0.94 L$. Vertical black dashed lines indicate integer values of $L_\mathrm{sd}/d_c$.}\label{fig:figS3}.
\end{center}
\end{figure}

The MC and FV schemes solve the same semiclassical Boltzmann equation (Eq.~\ref{eq:boltzmann}), yet via different numerical schemes. The MC scheme analytically computes the intersections of individual carrier trajectories with the device boundaries, while the FV scheme evolves the carrier distribution over the entire device as a function of time. Each scheme has its own advantages - the MC scheme is inherently faster because calculations are performed only at device boundaries, while the FV scheme allows for visualization of properties in the bulk of the device (such as current vortices). Figure~\ref{fig:figS3} compares $R_\mathrm{nl}$ (in arbitrary units) vs $L_\mathrm{sd}/d_c$ obtained using the two schemes at detector location $L_c = 0.1L$ in a square device of side $L$, with $L_\mathrm{wall} = 0.02L$, $w = 0.02L$ and $L_\mathrm{sd} = 0.94 L$. The results have been normalized to aid comparison, and indicate that both numerical schemes yield consistent results.  

\subsection{Appendix C: Effect of source and drain PC width $w$ on fractional peaks}

\begin{figure}[!htbp]
\begin{center}
\includegraphics[width=15cm]{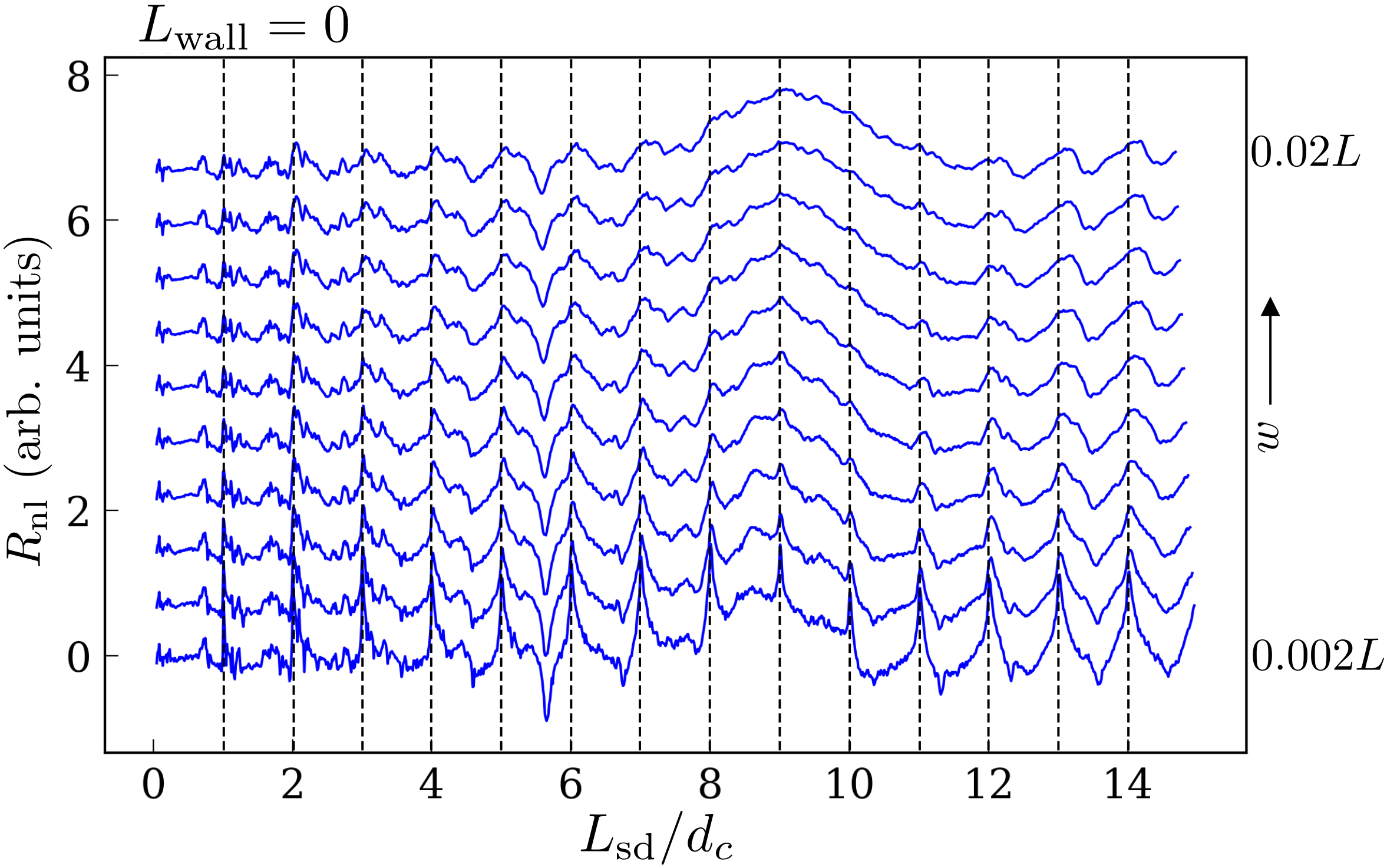}
\caption{$R_\mathrm{nl}$ vs $L_\mathrm{sd}/d_c$ simulated using the MC scheme for a square device of size $L \times L$ with varying source and drain width $w$ while $L_\mathrm{wall} = 0$ is kept constant. Widths $w$ are increased from 0.002$L$ to 0.02$L$ upwards. Vertical black dashed lines mark the locations of fractional peaks (integer values of $L_\mathrm{sd}/d_c$). The fractional peaks decrease in prominence as $w$ increases.}\label{fig:figS1}. 
\end{center}
\end{figure}

We observe that in the modeling in Fig.~\ref{fig:fig2}(b) with narrow $w = 0.002L$, the modeled oscillation amplitude of $R_\mathrm{nl}$ for the fractional peaks is larger than for the integer peaks, while the experimentally measured amplitude in Fig.~\ref{fig:fig1}(c-f) (where $w \approx 0.04L$) shows the opposite trend. Hence, using the MC scheme we modeled $R_\mathrm{nl}$ for variable source and drain PC width $w$, keeping $L_\mathrm{wall} = 0$. We use the same earlier square device, with dimensions $L \times L$ in arbitrary units and specular boundaries. Results are depicted in Fig.~\ref{fig:figS1}, showing that an increase in $w$ results in a decrease in fractional peak amplitudes and in a small systematic shift of the peak to $L_\mathrm{sd}/d_c \gtrsim n$. As $w$ is increased, the fractional peaks lose prominence and eventually the amplitudes of the integer peaks surpass amplitudes of the fractional peaks. Moreover, for $w = 0.002 L$ fractional peaks occur at exactly $d_c=L_\mathrm{sd}/n$ and are very sharp, much narrower than the integer TMF peaks, which are broadened by the angular spread of the injected carriers due to the existence of a Fermi contour. 

\subsection{Appendix D: Effect of distance of source and drain PCs from side walls $L_\mathrm{wall}$ on fractional peaks}

\begin{figure}[!htbp]
\begin{center}
\includegraphics[width=15cm]{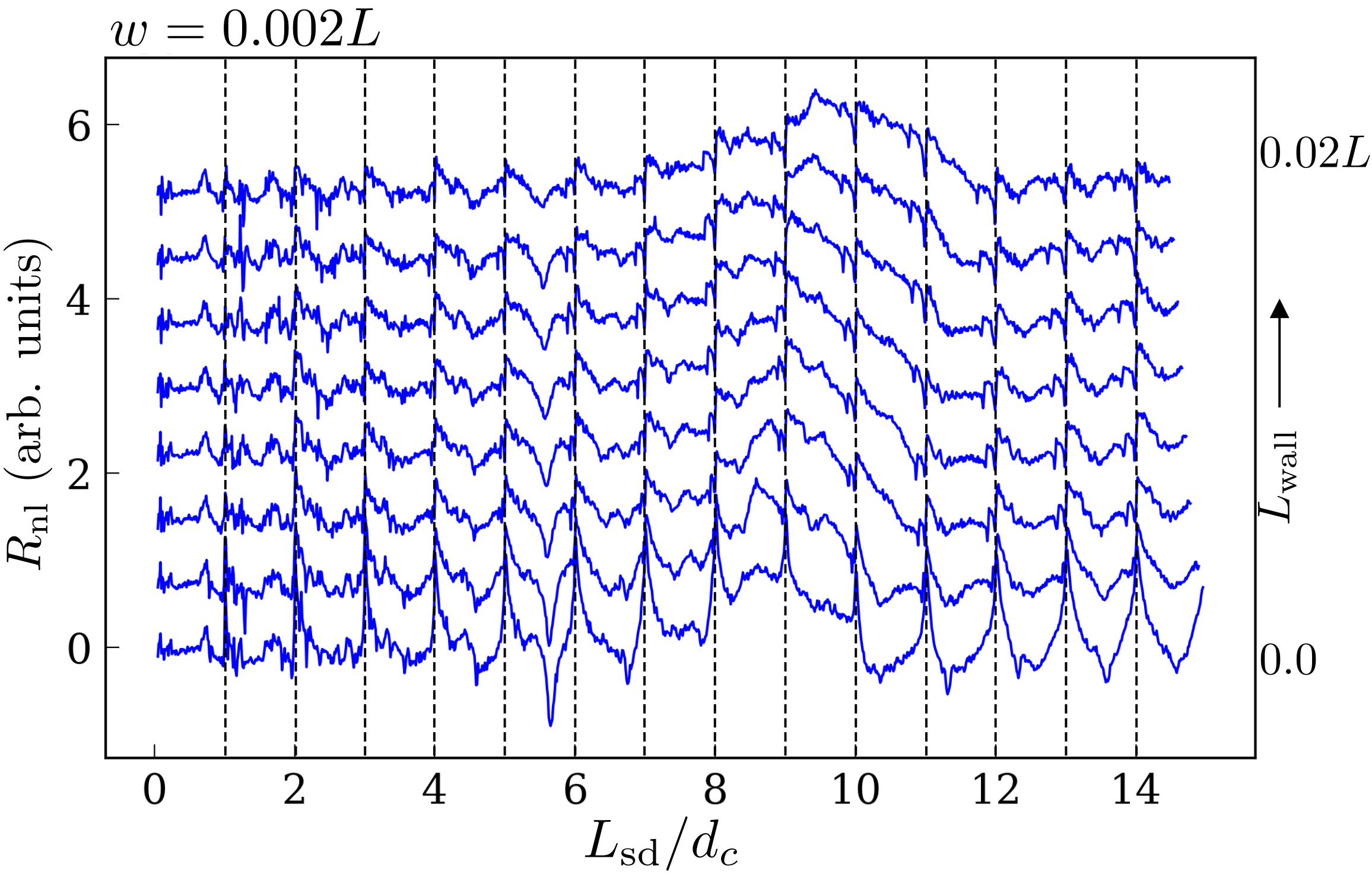}
\caption{$R_\mathrm{nl}$ vs $L_\mathrm{sd}/d_c$ simulated using the MC scheme for a square device of size $L \times L$ with varying $L_\mathrm{wall}$, while source and drain widths $w = 0.002L$ are kept constant. $L_\mathrm{wall}$ is increased from 0 to $0.02L$ upwards. Vertical black dashed lines mark the locations of fractional peaks (integer values of $L_\mathrm{sd}/d_c$). The fractional peaks decrease in prominence as $L_\mathrm{wall}$ increases.}\label{fig:figS2}.
\end{center}
\end{figure}

We identify the source and drain PC separation $L_\mathrm{wall}$ from the sidewalls as another factor that affects the amplitude of fractional peaks relative to that of the integer TMF peaks. In this section, we consider the evolution of $R_\mathrm{nl}$ when the source and drain PCs are situated at progressively larger distances $L_\mathrm{wall}$ from the sidewalls. Using the MC scheme we modeled $R_\mathrm{nl}$ for variable $L_\mathrm{wall} = 0.0 - 0.02L$, keeping fixed $w = 0.002L$. We use the same earlier square device, with dimensions $L \times L$ in arbitrary units and specular boundaries. Results are depicted in Fig.~\ref{fig:figS2}. For larger $L_\mathrm{wall}$, the fractional peaks are diffused and barely visible over the background in $R_\mathrm{nl}$. Evidently small $L_\mathrm{wall}$ favors robust fractional peaks. 

The effects of larger $w$ and $L_\mathrm{wall}$ are observed in the experimental results in Fig.~\ref{fig:fig2}(b) (where $w \approx 0.04 L$, $L_\mathrm{wall} \approx 0.01L$) and in the computational results in Fig.~\ref{fig:fig4}(f) (where $w \approx 0.02 L$, $L_\mathrm{wall} \approx 0.02L$). The fractional peak amplitudes are in these figures smaller than the integer TMF peak amplitudes (indicated by $n_c = 1$), and the fractional peaks are shifted to $L_\mathrm{sd}/d_c \gtrsim n$. 

\subsection{Appendix E: Quantum-coherent transport model}

While fractional peaks appear in simulations based on the semiclassical Boltzmann equation (Eq.~\ref{eq:boltzmann}) solved in either the FV or MC scheme, we here ask what the effect is of \emph{quantum-coherent} transport and how similar results are to semiclassical transport. The key differences when compared to the semiclassical model are the presence of quantum interference and a finite number $N_m$ of transverse modes injected through the PCs. To quantitatively address the comparison, we solve for ballistic quantum-coherent transport using the \texttt{KWANT} code \cite{Groth2014}, which solves the tight-binding model, 
\begin{align}
    H & = \sum_{i}4t\ket{i}\bra{i} - \sum_{i,j}t\exp\left(\frac{i}{2}\phi(x_i - x_j)(y_i + y_j)\right)\ket{i}\bra{j} \label{eq:Hamiltonian}
\end{align}
The above is obtained from a finite-difference approximation of the continuous Hamiltonian $H = \left(\mathbf{p} - e\mathbf{A} \right)^2/(2m)$ on a square lattice with spacing $a$, and in the Landau gauge with vector potential  $\mathbf{A} = (-By, 0)$. Here $t = \hbar^2/(2ma^2)$ and we set $a = \hbar = e = m = 1$. The dependence on $B$ enters through the non-dimensional parameter $\phi = Ba^2/\Phi_0$ (ratio of the magnetic flux through one cell of the square lattice to the flux quantum $\Phi_0 = \hbar/e$). The first term denotes the on-site energy, with a summation over all individual sites, and the second term a summation over nearest neighboring sites, with $x_i$ denoting the $x$ coordinate of the $i$ site (and similarly for $x_j,y_i,y_j$). The semiclassical model corresponds to the limit $N_m \gg 1$, whereas quantum-coherent transport has only been solved for in the few mode limit $N_m \ll 10$ (e.g., \cite{Stegmann, LaGasse, NanoResLett2022-17-Cole}). To approach the semiclassical limit using a tight-binding model for the given geometry is challenging since we require $L \gg w$ \emph{and} $w \gg \lambda_F$; the first inequality corresponds to the ideal condition needed for the fractional peaks to manifest, and the second condition signifies $N_m \gg 1$. We thus consider a 5000$\times$5000 site model with infinite leads placed on the bottom edge, each $100$ sites wide. We set $k_F = 0.31\pi/a$, consistent with the requirement needed for a continuous approximation ($k_F \ll \pi/a$) and a nearly circular Fermi contour. With these parameters, we obtain 32 modes. 
We compute the carrier densities $\rho_i$ at site $i$, and currents $J_{ij}$ from site $j$ to site $i$ using,
\begin{align}
    \rho_i & = \sum_{\alpha} \psi_{\alpha i}^* \psi_{\alpha i} \\
    J_{ij} & = i\sum_{\alpha} \left(\psi^*_{\alpha i} H_{ij} \psi_{\alpha j} - \psi_{\alpha i} H_{ij} \psi_{nj} \right)
\end{align}
where $\psi_{\alpha i}$ denote the modes originating from a chosen lead (either source or drain). The index $\alpha$ runs over all the modes up to energy $E_F$ in the lead and $i,j$ denotes the sites. $H_{ij}$ is the hopping matrix element from site $j$ to $i$, defined by $H = H_{ij}\ket{i}\bra{i}$, where $H$ is the Hamiltonian given in Eq.~\ref{eq:Hamiltonian}. The electron-hole symmetry at the Fermi surface allows for transport both by holes injected through the source PC and by electrons injected through the drain PC. Therefore, we need to compute scattering states originating from both the source and drain, and then evaluate the corresponding $\rho^h_i$ and $\rho^e_i$ for hole and electron modes respectively. The net carrier density is then $\rho_i = \rho^h_i - \rho^e_i$, which is proportional to the voltage at site $i$. The calculations for $J_{ij}$ follow similarly. 

\begin{figure}[!htbp]
\begin{center}
\includegraphics[width=17cm]{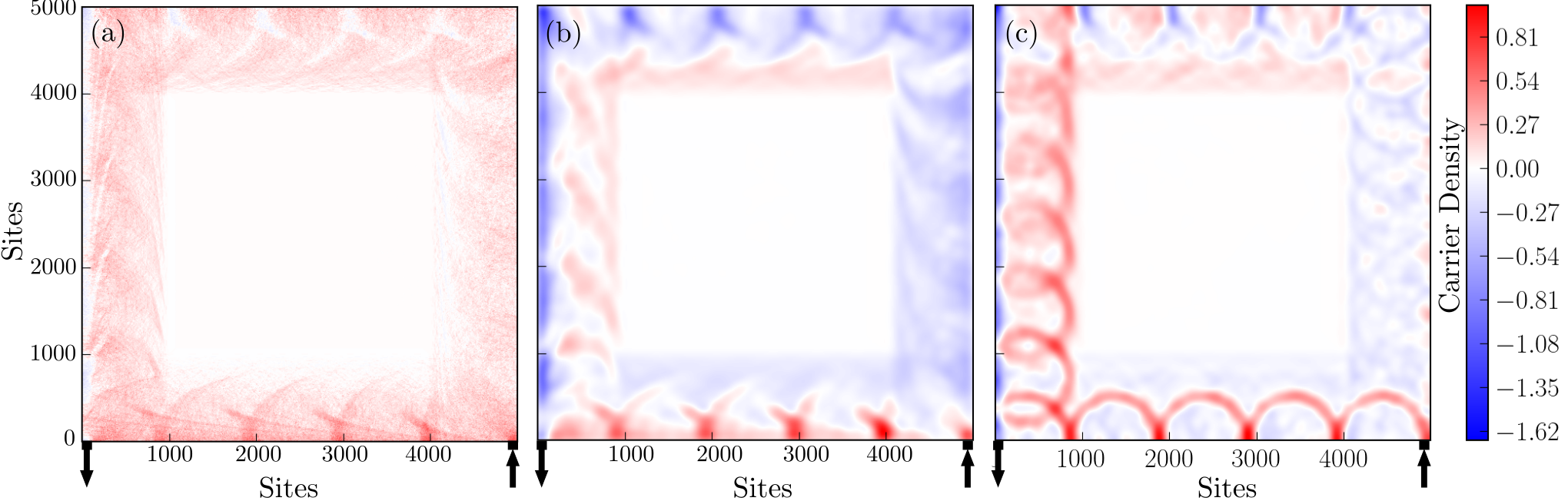}
\caption{Net carrier density ($\propto$ voltage) contour plots obtained using \texttt{KWANT} (quantum-coherent model) in a 5000$\times$5000 site device with $w = 100$ sites wide (corresponding to $w = 0.02 L$), $L_\mathrm{wall} = 0$ and $L_\mathrm{sd} = 0.98 L$, for the fractional $n = L_\mathrm{sd}/d_c = 5$ peak. (a) Net carrier density ($\propto$ voltage) obtained from \texttt{KWANT} with 32 injected hole and electron modes and by performing a summation over all modes. (b) Voltage averaged over $\sim50$ adjacent sites and by performing a summation over the 32 injected hole and electron modes. (c) Voltage averaged over $\sim50$ adjacent sites and obtained by injecting only single hole and electron modes.\label{fig:figS5}}
\end{center}
\end{figure}

Figure~\ref{fig:figS5}(a) shows the net carrier density ($\propto$ voltage) at every lattice site obtained from \texttt{KWANT} for the fractional $n = L_\mathrm{sd}/d_c = 5$ peak for a 5000$\times$5000 site, with $w = 100$ sites wide (corresponding to $w = 0.02 L$), $L_\mathrm{wall} = 0$, $L_\mathrm{sd} = 0.98 L$, and with 32 hole modes injected through the source PC and 32 electron modes injected through the drain PC. In order to compare the voltages and currents obtained from the tight-binding \texttt{KWANT} model to the semiclassical model, the tight-binding solutions need to be appropriately averaged. Figure~\ref{fig:figS5}(b) shows the density (as shown in Fig.~\ref{fig:figS5}(a)) averaged over $\sim50$ adjacent sites, corresponding to the low-energy limit of the lattice solution, which now resembles the semiclassical model.

\begin{figure}[!htbp]
\begin{center}
\includegraphics[width=15cm]{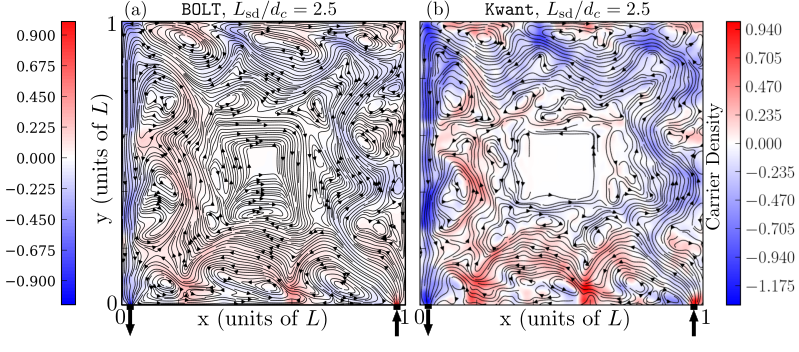}
\caption{Comparison of current streamlines and carrier density (proportional to voltage) contour plots for $L_\mathrm{sd}/d_c = 2.5$ in a square device with dimensions $L \times L$, $L_\mathrm{wall} = 0.02L$, $w = 0.02L$ and $L_\mathrm{sd} = 0.94 L$ obtained using (a) \texttt{BOLT} (semiclassical transport in FV scheme) and (b) \texttt{KWANT} (quantum-coherent model with density summed over 32 injected hole and electron modes and spatially averaged over $\sim50$ adjacent sites).}\label{fig:figS4}.
\end{center}
\end{figure}

\begin{figure}[!htbp]
\begin{center}
\includegraphics[width=15cm]{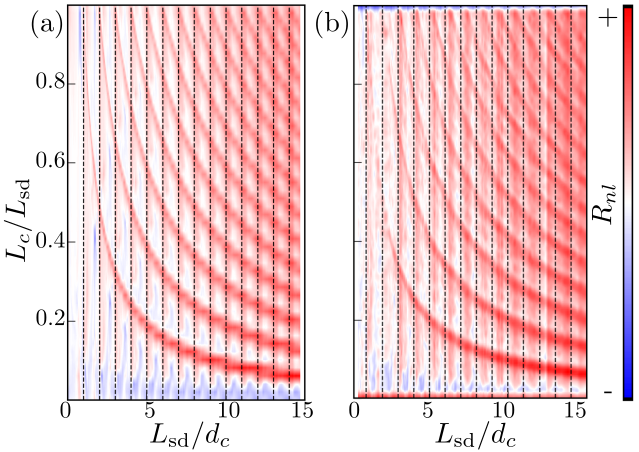}
\caption{Contour plot of simulated $R_{nl}$ vs $L_c/L_\mathrm{sd}$ and $L_\mathrm{sd}/d_c$ in a square device with dimensions $L \times L$, $L_\mathrm{wall} = 0$, $w = 0.02L$, $L_\mathrm{sd} = 0.98 L$ and specular device boundaries, obtained using the (a) semiclassical model (in MC scheme) and (b) quantum-coherent model with density summed over 32 injected hole and electron modes and spatially averaged over $\sim50$ adjacent sites). Vertical black dashed lines indicate integer values of $L_\mathrm{sd}/d_c$.}\label{fig:figS7}.
\end{center}
\end{figure}

The question arises whether the fractional peaks also appear if just a few modes are considered or whether they require a summation of several modes. Figure~\ref{fig:figS5}(c) shows the voltage contour plot (averaged over $\sim50$ adjacent sites) when only a single hole mode and a single electron mode are considered as compared to Fig.~\ref{fig:figS5}(b) where 32 modes of each are considered. In our color convention, the single-mode calculation shows alternating positive (red) and negative (blue) voltages at the bottom edge of the device, while in the multimode calculation, the voltage at the bottom edge remains positive (red) throughout. The results of the multimode calculation are consistent with the observation of the fractional peak - a hole overdensity everywhere on the bottom edge, while the single-mode calculation does not indicate this observation. Another noticeable effect in both voltage contour plots (but more so in the single-mode) is that the cyclotron orbits (red semicircular structures) do not impinge on the drain PC at exactly the fifth reflection from the boundary, instead appearing to land somewhat farther. This can be attributed to the effect of finite $w$, with the number of reflections compounding the effect further (as observed in \cite{gupta2021NatComm}). Finally, in the multimode calculation the intensity of red semicircular orbits decays as we move farther away from the source PC. The single-mode calculation, however, fails to capture this important effect. Figure~\ref{fig:figS4} directly compares the densities and currents between \texttt{KWANT} and \texttt{BOLT} (FV scheme) at $L_\mathrm{sd}/d_c = 2.5$, in the same geometry as shown in Figs.~\ref{fig:fig4}(a-d) of the main text. The qualitative match is evident. Figure~\ref{fig:figS7} compares the contour plots of $R_{nl}$ vs $L_c/L_\mathrm{sd}$ and $L_\mathrm{sd}/d_c$ between the semiclassical transport model (using the MC scheme) and the quantum-coherent model for a square device with specular boundaries, length $L$, $L_\mathrm{wall} = 0$, $w = 0.02L$ and $L_\mathrm{sd} = 0.98 L$. Fractional peaks are clearly evident in both the semiclassical and quantum-coherent transport models, with the location of peaks matching closely at lower $L_\mathrm{sd}/d_c$ values. 

\subsection{Appendix F: ARTS plots at integer $L_\mathrm{sd}/d_c$}

\begin{figure}[!htbp]
\begin{center}
\includegraphics[width=16cm]{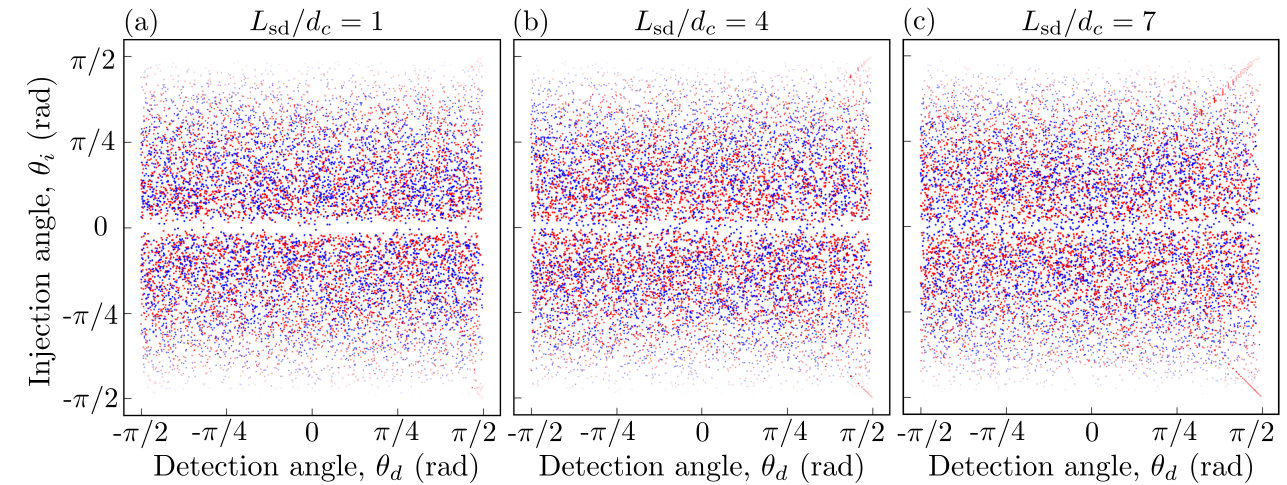}
\caption{Comparison of ARTS plots in a $L \times L$ device, with $w = 0.002 L$, $L_\mathrm{wall} = 0$, $L_\mathrm{sd} = 0.998 L$ and $L_c = 0.1L$, obtained using the MC scheme for fractional peaks at (a) $L_\mathrm{sd}/d_c = 1$ (b) $L_\mathrm{sd}/d_c = 4$ and (c) $L_\mathrm{sd}/d_c = 7$.}\label{fig:figS6}.
\end{center}
\end{figure}

As illustrated in Figs.~\ref{fig:fig3}(f,h) at the condition for appearance of fractional peaks, i.e. at integer $n = L_\mathrm{sd}/d_c$, the ARTS plots show a characteristic horizontal gap around $\theta_d \approx 0$. This gap indicates a lack of trajectories that reach the detector PC. We show in Fig.~\ref{fig:figS6} that this gap blurs with increasing $n = L_\mathrm{sd}/d_c$. As $n$ increases, the trajectories traverse longer path lengths before reaching the source/drain, increasing their likelihood of being randomized. The randomization of the trajectories leads to an increased probability that they reach the detector PC, instead of the source/drain PCs. Note that, as mentioned in the main text, the randomization here is deterministic; each trajectory follows an analytic solution obtained by a specular reflection off the device boundaries. 

\FloatBarrier

\newpage
\setcounter{page}{1}
\renewcommand\refname{References\\[0.1cm]}

\end{document}